\documentclass[aps,prl,reprint]{revtex4-2}

\usepackage{amsmath}
\usepackage{amssymb}
\usepackage{amsthm}
\usepackage{bm}
\usepackage{bbm}
\usepackage{graphicx}
\usepackage{xcolor}
\usepackage{physics}

\usepackage{hyperref}
\begin{document}

\title{Non-Markovian dissipation as a resource for quantum reservoir computing}

\author{Luca Nigro}
\author{Francesco Monzani}
\author{Enrico Prati}
\email{enrico.prati@unimi.it}
\affiliation{Università degli Studi di Milano, Department of Physics, Via Celoria 16, 20133 Milano, Italy}

\date{\today}

\begin{abstract}
The control of open-system dynamics provides a powerful mechanism for using quantum information to process sequential tasks.
While quantum reservoir computing typically relies on Markovian dissipation to process sequential data, the computational role of non-Markovian memory effects remains largely unexplored. 
We introduce a framework for quantum reservoir computing where non-Markovianity is explicitly modeled and regulated using fractional derivatives. By employing fractional time subordination, we generate tunable, heavy-tailed relaxation dynamics that govern the information backflow between the system and its environment.
%
Non-Markovian information backflow regulates the overall temporal retention of the system, maximizing short-term linear memory capacity .
Within an optimal operating regime, fractional non-Markovianity redistributes the reservoir memory towards recent inputs, improving both linear memory capacity and nonlinear prediction accuracy at short delays, at the expense of long-range temporal retention.
System-environment interaction proves being not merely a requirement for quantum reservoir computing, but an active resource that embeds memory retention directly into the quantum evolution.
\end{abstract}

\maketitle
\textit{Introduction}---The dynamics of open quantum systems can be non-Markovian, typically if the system-environment coupling induces memory effects on a timescale comparable to that of the evolution of the system itself~\cite{breuer2002theory,tarasov2021,vacchini_book, mandarino}.
Understanding such memory effects is not only a fundamental physics problem but an urgent practical challenge, as they induce measurable consequences~\cite{rivas2014quantum, devega2017} for quantum technologies unavoidably operating in strongly correlated, non-Markovian environments~\cite{agarwal2024, perego2026, haase2018, breuer2016colloquium}.

Dissipation has already proved beneficial to quantum information processing~\cite{verstraete2009quantum}, as engineered dissipative dynamics can drive systems into target quantum states~\cite{non_uni_prx, non-uni.barren, sannia2024engineered} and open-system effects can assist certain computational tasks~\cite{diehl2008quantum}.
Yet, the effects of non-Markovian dynamics in temporal information processing are not widely utilized, with only a few studies present in the literature \cite{bylicka2013,Zambon2025,Porto2025,goswami2025,Berk2021,Berk2023,laine2012}. Nevertheless, these dynamics can serve as a powerful computational resource, as they naturally encode long-range temporal correlations.

To systematically address the latter question, we employ quantum reservoir computing (QRC)~\cite{jaeger2001echo, maass2002liquid, fujii2017quantum, nakajima2019quantum, mujal2021opportunities, chen2019learning, chen2020temporal, ghosh2019quantum, govia2021nonlinear, martinez2021dynamical, nokkala2021gaussian, suzuki2022natural, mujal2023timeseries, hu2024overcoming, sannia2024dissipation, kobayashi2026edge}, a versatile temporal processing paradigm applicable across diverse platforms~\cite{rc.naka, monz_uni, tanaka2019recent, grollier, senanian2024microwave, franceschetto, ghosh2021realising, gotting2023exploring, ph1, ph2, neutral.atoms, boso1, nokkala2021gaussian, ghosh2019quantum, Spagnolo2022, pfeffer, magri, dudas2023quantum, yama}. By utilizing fixed quantum dynamics with a simple linear readout~\cite{nakajima2020physical,fujii2021chapter}, QRC serves as an ideal testbed to link computational performance directly to underlying physics~\cite{gotting2023,kora2024,palacios2024,xia2023,kutvonen2020,martinez2020ipc,xiong2025role}. Notably, dissipation acts as a resource here, where tunable non-unital noise provides the fading memory and expressivity required for universality~\cite{sannia2024dissipation, domingo2023taking, kubota2023temporal, monzani2024nonunital, ricci2025damping}. This framework is also well suited to investigate non-Markovianity~\cite{sannia2025nonmarkov, sasaki2025hamiltoniandrivenarchitecturesnonmarkovianquantum}, offering a path to harness memory effects as a computational resource.

In this Letter, we study a spin-chain quantum reservoir~\cite{fujii2017quantum} subject to non-Markovian open-system dynamics, modeled by a generalized Gorini--Kossakowski--Sudarshan--Lindblad (GKSL) master equation involving a fractional time derivative, which encodes power-law memory kernels and allows continuous interpolation between Markovian and strongly non-Markovian regimes~\cite{tarasov2008fractional, iomin2009, wei2023, tarasov2021, laskin2000fractional,peng2026, Metzler2000, peng2026}. First, we rigorously quantify non-Markovianity via distinguishability and divisibility metrics~\cite{demir2025, Vacchini2011, breuer2016colloquium}. Next, we establish the theoretical conditions for a universal reservoir computer~\cite{monz_uni,maass, sannia2024dissipation,nokkala2021gaussian,chen2019learning,chen2020temporal}, namely the echo state property, fading memory, and input separability under non-Markovian dissipation. The performance of such a system is benchmarked over linear and nonlinear memory tasks~\cite{dambre2012, martinez2020ipc, mujal2021opportunities}. We show that non-Markovianity increases short-term memory capacity by a factor of three for linear processing, while reducing prediction errors in nonlinear tasks by up to an order of magnitude. 
Our findings for the specific case studied here suggest that engineered non-Markovian memory can provide a distinct computational advantage, offering a promising framework for exploring memory effects across physical platforms relevant to near-term quantum hardware~\cite{agarwal2024, perego2026, haase2018}.

\textit{Basics of Reservoir computing}---Reservoir computing (RC) is a
machine learning paradigm for sequential data, in which a temporal input signal $\{u_k\}$ drives a fixed dynamical system, namely the reservoir, whose internal state is read out by a simple trainable map~\cite{jaeger2001echo,maass2002liquid}.
Notably, the reservoir is not optimized. It acts as a nonlinear temporal filter that expands the input into a high-dimensional feature space, where a linear readout suffices to approximate a broad class of target functionals. Training is therefore reduced to a convex regression problem, which makes RC particularly attractive for physical substrates whose internal
parameters are not individually addressable, as it may be the case for quantum hardware.
Because the readout is memoryless, all the temporal processing power resides in the reservoir dynamics. The state at step $k$ must retain a usable trace of the past inputs $u_{k-1}, u_{k-2}, \dots$, while remaining independent of how
the system was initialized. These two requirements are formalized by the echo state property (ESP) and by fading memory~\cite{jaeger2001echo,boyd.chua}. The ESP
demands that the reservoir state be asymptotically a function of the input sequence alone, so that any dependence on the initial condition is washed out.
On the other hands, fading memory ensures that the influence of an input decays with its distance in the past, so that recent inputs dominate the current state. Together they
guarantee that the input--output map is well defined and continuous, which is
the basis of the universality results for reservoir
computing~\cite{chen2019learning, nokkala2021gaussian, sannia2024dissipation, monzani2024nonunital}. In particular, a dissipative mechanism is constitutive, since it induce the contraction of the state space, which ensure these properties to hold \cite{sannia2024dissipation, hu2024overcoming, monzani2024nonunital, kubota2023temporal, chen2019learning, chen2020temporal}.
Thus, characterizing a reservoir therefore amounts to resolving how its memory is distributed over past inputs. For this, we employ two complementary benchmarks well-rooted in RC,, namely short-term memory (STM) and NARMA task ~\cite{dambre2012}. The STM
capacity quantifies how faithfully a linear readout reconstructs a single delayed input $u_{k-\tau}$, and thus describes the linear memory profile. The NARMA-$p$ task, conversely, targets an output that depends nonlinearly on $p$ past inputs and outputs~\cite{narma}, probing the joint nonlinear processing of an extended history.

\textit{Quantum reservoir computing model} --- The reservoir is modeled as a one-dimensional spin-$\frac{1}{2}$ chain 
of $N$ spins. Such a system represents a well-established embodiment of quantum reservoir computing~\cite{fujii2017quantum, 
sannia2024dissipation, martinez2021dynamical, martinez2020ipc}, whose simplicity and analytical tractability make it an ideal testbed for the manifestation of the computational consequences of non-Markovian dynamics. Its dynamics is governed by the general quantum Ising-Lenz Hamiltonian over a one-dimensional lattice
\begin{equation}
    H_{\mathrm{res}}= -J\sum_{i=1}^{N-1}\sigma_i^z\sigma_{i+1}^z 
    - \sum_{i=1}^{N}h_i\,\sigma_i^x \,,
    \label{eq:hamiltonian_ising}
\end{equation}
where $J$ is the spin-spin coupling factor and the coefficients $h_i$ represent the local transverse field. Throughout the work, we set $J=1.0$ and $h_i = 0.2$ for all $i=1,\dots,N$. These values are chosen based on the phase diagram analysis by Martinez-Peña \textit{et al.}~\cite{martinez2021dynamical}, which identifies this specific regime as the optimal operating region for maximizing the information processing capabilities of the reservoir while avoiding trivial or overly chaotic dynamics. To construct an input-output mapping, an input signal $\{u_k\}_{k=1}^T \in [0,1]^T$ is encoded in the reservoir by modulating the local transverse field. Namely, each value $u_k$ is encoded in the input-dependent Hamiltonian, which leads the evolution for a time interval $\Delta t$, as
\begin{equation}
    H_{\mathrm{input}}(u_k) = 
     \sum_{i=1}^{N}\left(g_i\, u_k\right)\sigma_i^x,
    \label{eq:hamiltonian_input}
\end{equation}
where $g_i$ are drawn from a uniform distribution $g_i \in [0, 0.5]$. We deliberately employ a unitary input encoding—rather than the state-preparation method~\cite{fujii2017quantum}—to ensure that the contractivity required for the echo state property and fading memory arises solely from the dissipative dynamics. In our model, the non-unitality of the amplitude damping channels, combined with the fractional memory kernel, serves as the unique mechanism for stabilizing the temporal processing of the reservoir, avoiding information loss due to repeated measurements \cite{ricci2025damping, hu2024overcoming}.
During the evolution, we apply the time-multiplexing strategy of virtual nodes~\cite{nakajima2019quantum, fujii2017quantum}, so
the expectation values of all local Pauli operators $\{\langle\sigma_i^\eta\rangle\}$, with $\eta\in \{x,y,z\}$ and $i = 1,\ldots,N$, are measured at $V = 8$ discrete times $t_v = v\,\Delta t/V$. Such a method expands the feature space from $3N$ to $3NV$ observables per time step, enriching the readout without increasing the physical system size. At each step $k$, these measurements form the state vector $\mathbf{x}_k \in \mathbb{R}^{3NV}$, which acts as input to a linear readout $\widehat{y}_k = \mathbf{W}\,\mathbf{x}_k$. 
The weight matrix $\mathbf{W}$ is the only trainable component of the model. 
It is computed via ridge regression by minimizing the mean squared error between the predicted output $\hat{y}_k$ and the target $y_k$ over the training set. 

\begin{figure}[tp]
    \centering
    \includegraphics[width=0.8\linewidth]{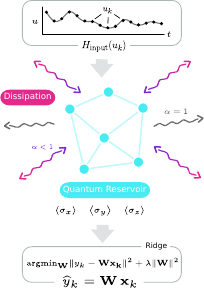}
    \caption{A sketch of the architecture of a quantum reservoir. The sequential input $i$ is injected through a unitary Hamiltonian, then processed by the spin system, and finally read out by the expected value of some observables. The linear readout is trained to optimize the desired task with a Ridge regression. The spin system undergoes Markovian (indicated by grey arrows, $\alpha=1$) and non-Markovian (purple arrows, $\alpha<1$) dissipation.}
    \label{fig:fig.0}
\end{figure}

\textit{Dissipative dynamics of the reservoir}---Dissipation plays a constructive role in quantum reservoir computing~\cite{sannia2024dissipation}, since it provides the non-unital dynamics required to ensure fading memory and separate distinct input sequences~\cite{kubota2023temporal,monzani2024nonunital}. We introduce dissipation via amplitude damping jump operators $L_i = \sigma_i^- = |0\rangle\langle 1|_i$ for $i = 1,\ldots,N$. A schematic representation of the model is shown in Fig.~\ref{fig:fig.0}.
The open-system dynamics of the reservoir density matrix $\rho$ are thus governed by the GKSL master equation~\cite{breuer2002theory}:
\begin{equation}
\frac{d\rho}{dt} = -i[H,\rho]+ \gamma\sum_{i=1}^{N}\left(\sigma_i^-\rho \sigma_i^+ - \frac{1}{2}\lbrace\sigma_i^+\sigma_i^-, \rho\rbrace\right)
\label{eq:lindblad}
\end{equation}
where $\gamma > 0$ is the uniform dissipation rate, and, for each time-step $\Delta t$, the driven Hamiltonian is given by $H = H_{\mathrm{res}} + H_{\mathrm{input}}(u_k)$. 
The entire temporal evolution of the reservoir, from input encoding to readout, is strictly governed by Eq.~\eqref{eq:lindblad}. 
This equation constitutes the central dynamical object of our work and serves as the starting point for the non-Markovian generalization introduced below.

\textit{Non-Markovianity of fractional evolution}---To inject controllable memory effects into the reservoir, we generalize the open-system dynamics by replacing the standard first-order time derivative in Eq.~\eqref{eq:lindblad} with a Caputo fractional derivative of order $\alpha \in (0,1]$.  
The Caputo definition is adopted as it allows for standard initial conditions, making it physically consistent with open-system state evolution. Indeed, fractional calculus provides the fundamental mathematical framework for describing systems with anomalous transport and complex memory, such as sub-diffusion in porous media and viscoelastic materials \cite{Metzler2000} and open quantum systems~\cite{peng2026}. For such systems, the dynamics is governed by heavy-tailed trapping events, meaning the probability of exceptionally long wait times remains non-negligible. Similarly, when subjected to strong environmental correlations, a quantum register undergoes anomalous relaxation that retains long-range temporal correlations, a behavior precisely captured by fractional dynamics, since it provides a macroscopic dial to tune the non-Markovian memory directly through the evolution equations.\\
We rigorously solve the resulting integro-differential equation utilizing the mathematically exact framework of time subordination~\cite{Stanislavsky2008, demir2025}. This paradigm expresses the state $\rho(t)$ in physical time as a continuous integral over a standard memoryless trajectory $e^{\mathcal{L}s}\rho(0)$, where $\mathcal{L}$ is the GKSL generator. The dynamics is governed by an internal operational time $s$, namely
\begin{equation}\label{eq:subord}
    \rho(t) = \int_{0}^{\infty} f_{\alpha}(s, t)\, e^{\mathcal{L}s}\rho(0) \dd{s}.
\end{equation}
Crucially, the stochastic kernel $f_{\alpha}(s, t)$ is the probability density of an inverse one-sided Lévy stable subordinator (see Supplementary Material for details). Physically, this unravels the complex memory convolution into a randomized operational clock. It allows us to simulate strongly non-Markovian dynamics using Monte Carlo trajectories, natively capturing heavy-tailed algebraic decay without relying on heuristic approximations (see Supplementary Material for details). Since the driven Hamiltonian is constant within each input interval, Eq.~\eqref{eq:subord} is applied on each interval with generator $\mathcal{L}(u_k)$ and the state inherited from the previous step. The fractional memory is therefore re-anchored at every reservoir update rather than propagated across the entire input sequence.
\begin{figure}[tp]
\centering
\includegraphics[width=\linewidth]{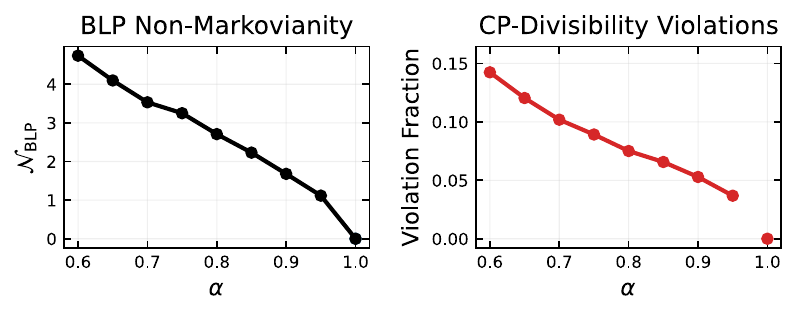}
\caption{Validation of non-Markovian dynamics driven by the fractional order $\alpha$. The left panel shows the Breuer-Laine-Piilo (BLP) measure $\mathcal{N}_{\text{BLP}}$, and the right panel displays the CP-divisibility violation fraction. Solid lines denote the maximum observed value for the BLP measure and the mean value for the CP-divisibility fraction across $50$ randomly sampled initial state pairs. Both metrics vanish in the Markovian limit ($\alpha=1$) and monotonically increase in strongly fractional regimes ($\alpha < 1$), confirming the onset of power-law memory effects. Results are computed for a single-spin system with decay rate $\gamma = 0.4$, time step $\Delta t = 0.1$, and $100$ integration steps.}
\label{fig:non_markov}
\end{figure}
\noindent As a first result, we verify that the fractional driver successfully breaks the Markovianity of the system and induces quantum memory effects. To quantify this, we evaluate two established criteria, namely the Breuer-Laine-Piilo (BLP) measure of information flow, $\mathcal{N}_{\text{BLP}}$~\cite{breuer2009}, and the violation of completely positive (CP) divisibility \cite{cpviolation}. The BLP measure detects the backflow of information from the environment to the system by tracking temporary increases in the trace distance between two initially distinct quantum states $D(\rho_1, \rho_2) = \norm{\rho_1 - \rho_2} /2$, where $\norm{A}= \Tr[\sqrt{A^\dag A}]$ is the trace norm.
Namely, for a dynamical map $\Phi$ that dictates the evolution for the states $\rho_1$ and $\rho_2$, the definition of the BLP measure is
\begin{equation}
    \mathcal{N}_\text{BLP} [\Phi] = \max_{\rho_1, \rho_2}\int_{\sigma>0} \sigma(\rho_1, \rho_2; t) \dd{t}
\end{equation}
where $\sigma(\rho_1, \rho_2; t) = \frac{\dd}{\dd{t}} D(\rho_1, \rho_2)$ is the rate of change of the distance. 
Conversely, the CP-divisibility violation signals the breakdown of memoryless quantum dynamical maps. Namely, we report the violation fraction defined as the proportion of time-ordered pairs
$(t_i,t_j)$ with $i<j$ for which $D\left(\rho_1(t_j),\rho_2(t_j)\right)>D\left(\rho_1(t_i),\rho_2(t_i)\right)+\varepsilon$, out of the
$N(N-1)/2$ such pairs on the sampled grid, with $\varepsilon=10^{-10}$ a
numerical tolerance. It measures
the fraction of the evolution affected by backflow rather than the number
of revivals, and is therefore complementary to $\mathcal{N}_{\text{BLP}}$,
which weights their magnitude. Both are evaluated per pair of initial
states and averaged over the $50$ sampled pairs. Fig.~\ref{fig:non_markov} illustrates both metrics as a function of the fractional order $\alpha$. In the standard Markovian limit ($\alpha = 1$), the environment is strictly memoryless. Consequently, the BLP measure vanishes identically, and the map remains strictly CP-divisible. However, as the fractional driver is activated by decreasing $\alpha$, the system undergoes a clear transition towards non-Markovianity. Indeed, we observe a strictly monotonic increase in both the BLP measure and the CP-divisibility violation fraction. The robust $\mathcal{N}_{\text{BLP}} > 0$ signal confirms a sustained backflow of information continuously feeding past dynamics into the current state. This establishes $\alpha$ as a precise control dial for tuning the non-Markovianity of the reservoir.

\textit{Non-Markovian temporal processing}---A reliable reservoir computer must process incoming data while maintaining ESP, which consists of an evolution driven by the input regardless of the initial condition of the reservoir, and fading memory, namely the ability to progressively forget past input values \cite{boyd.chua, tigher.esp}.
\begin{figure}[tp]
    \centering
    \includegraphics[width=0.65\linewidth]{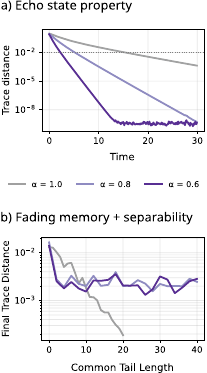}
    \caption{Verification of the Echo State Property (ESP) and fading memory. (a) Convergence of distinct initial conditions. Two nearly orthogonal random states (initial trace distance $\approx 1$) are driven by the same random input sequence. Over time, the trace distance vanishes, demonstrating that the system asymptotically forgets its initial state. The dynamics are shown for Markovian ($\alpha=1$, gray) and fractional ($\alpha=0.8$, light violet; $\alpha=0.6$, dark violet) regimes. (b) Fading memory and input separability. A single initial state is driven by two distinct random input time series that share only their final $\ell$ elements. The trace distance between the two resulting states is tracked as a function of the common tail length $\ell$. Simulations are performed on a four-spin linear array with a decay rate $\gamma = 0.4$. The system is evolved for $300$ time steps with $dt = 0.1$, and the curves represent the mean over $30$ random realizations.}
    \label{fig:esp}
\end{figure}

\begin{figure*}[t]
    \centering
    \includegraphics[width=0.7\textwidth]{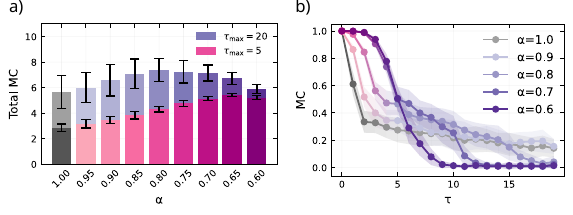}
    \caption{Impact of the fractional order $\alpha$ on short-term memory capacity. (a) Total memory capacity (MC) as a function of $\alpha$, with error bars representing the standard deviation over 20 independent iterations. The violet and pink bars denote the cumulative MC evaluated up to maximum delays of $\tau_{\max} = 20$ and $\tau_{\max} = 5$, respectively. The difference between these bounds isolates the capacity retention in the memory tails ($\tau > 5$). The Markovian reference limit ($\alpha = 1$) is indicated in gray. (b) Temporal decay profile of the memory capacity $\text{MC}(\tau)$ across discrete delays $\tau$, parameterized by varying $\alpha$. Line colors correspond to the $\alpha$ values defined in panel (a) Simulations are performed on a linear chain of 
$N=4$ spins with decay rate $\gamma = 0.4$.\label{fig:stm_task}}
\end{figure*}

\noindent To verify the ESP, we initialize two identical reservoirs with uniformly sampled amplitudes, resulting in approximately orthogonal initial states, so that their initial trace distance is maximal. We then let both density matrices evolve driven by an identical random input sequence. Fig.~\ref{fig:esp}(a) tracks the trace distance between the resulting states. This distance asymptotically vanishes for both the Markovian baseline ($\alpha=1$) and the fractional regimes ($\alpha < 1$). indicating that, even with a non-Markovian dynamics, the ESP is guaranteed. In particular, we notice that the fractional driver, which describes the non-Markovian evolution, slows the state-space contraction. However, it firmly preserves the strict contractivity required to wash out initial conditions.\\
The remaining requirements for a proper quantum reservoir are input separability and fading memory. To assess these properties as a function of $\alpha$, we drive a single initial state with two distinct random input sequences. These streams share only their final $\ell$ elements. Fig.~\ref{fig:esp}(b) plots the trace distance between the two output states against the common tail length $\ell$. The trace distance converges to zero as the shared recent history lengthens. The non-Markovian reservoir successfully separates distinct inputs while systematically erasing the distant past. These results provide evidence that the fractional open-system dynamics operate within a stable computational regime. Moreover, we notice that the Markovian dissipation ensures linear fading memory, continuously erasing past information while the input is encoded. 
The non-Markovian dissipation, on the other hand, initially leads to a loss of information on short timescales, but reaches a plateau in the final trace distance between the state of the two reservoirs around $\ell=5$ in the common tail. This peculiar behavior in short-term processing also reflects on the computational capabilities, as discussed in the following Sections. 

\textit{Short-term memory capacity}---We begin to quantify the computational capabilities of a quantum reservoir undergoing non-Markovian dissipation by testing its linear memory capacity. The Short-Term Memory (STM) capacity quantifies the ability of the reservoir to reconstruct past inputs at temporal delay $\tau$, and is measured as
\begin{equation}
    \text{MC}_\tau = \frac{\text{Cov}^2\left[y_k(\tau), \widehat{y}_k\right]}{\text{Var}\left[y_k(\tau)\right]\text{Var}\left[\widehat{y}_k\right]},
\end{equation}
where $y_k(\tau) = u_{k-\tau}$ is the target delayed input and $\widehat{y}_k$ is the predicted output. We benchmark the fractional reservoir by evaluating $\text{MC}_\tau$ as a function of $\tau$ for varying $\alpha$. Fig.~\ref{fig:stm_task} shows the impact of the fractional order on the memory retention profile. In particular, we present total MC at two different orders $\tau_{\mathrm{max}} = 5, 20$, defined as $\text{MC}^{\tau_{\mathrm{max}}} = \sum_{\tau = 0}^{\tau_{\mathrm{max}}}\text{MC}_{\tau}$ .\\
In the Markovian limit ($\alpha = 1$), the memory capacity decays exponentially with $\tau$, as expected from standard GKSL dynamics. Activating the fractional driver ($\alpha < 1$) fundamentally reshapes this temporal distribution. Non-Markovian dynamics substantially enhance the capacity for recent inputs ($\tau \lesssim 5$). Indeed, the fractional reservoir maintains a sharper and more distinguishable representation of its immediate history. However, this short-time enhancement comes with a structural trade-off, since the capacity in the memory tails ($\tau > 5$) drops below the Markovian baseline. The non-Markovian environment effectively concentrates the linear processing power of the reservoir on the immediate past, thereby reducing long-range temporal correlations. This is compatible with the plateau observed in the fading memory analysis, where non-Markovian dynamics were shown to retain a residual trace distance for large common tail lengths, indicating a reduced ability to forget distant past inputs, hindering the ability to process long-term inputs.

\begin{figure*}[tp]
    \centering
    \includegraphics[width=0.7\textwidth]{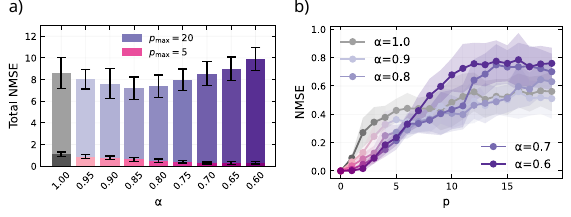}
    \caption{Impact of the fractional order $\alpha$ on NARMA task performance. (a) Total Normalized Mean Square Error (NMSE) as a function of $\alpha$, with error bars representing the standard deviation over 20 independent iterations. The violet and pink bars denote the cumulative NMSE evaluated up to maximum NARMA orders of $p_{\max} = 20$ and $p_{\max} = 5$, respectively. The difference between these bounds isolates the error accumulation associated with higher-order nonlinear dependencies ($p > 5$). The Markovian reference limit ($\alpha = 1$) is indicated in gray. (b) Error profile $\text{NMSE}(p)$ across discrete NARMA orders $p$, parameterized by varying $\alpha$. Line colors correspond to the $\alpha$ values defined in panel (a). Simulations are performed on a linear chain of 
$N=4$ spins with decay rate $\gamma = 0.4$. }\label{fig:narma_task}
\end{figure*}

\textit{Nonlinear temporal processing}---We assess the nonlinear processing capabilities using the NARMA-$p$ benchmark \cite{narma}, where the target output is defined as
\begin{equation}
    y_{k+1} = a y_k + b y_k \sum_{j=0}^{p-1} y_{k-j} + c u_{k-p+1} u_k + d.
\end{equation}
The specific parameters $\{a, b, c, d\}$ are chosen according to standard benchmark conventions~\cite{narma} to prevent unbounded trajectories ($y_k \to \infty$) due to non-linear feedback, while ensuring non-trivial long-range temporal dependencies.
Performance is measured by the Normalized Mean Square Error
\begin{equation}
    \text{NMSE} = \frac{\mathbb{E}\left[(y_k - \widehat{y}_k)^2 \right]}{\text{Var}\left[y_k\right]}.
\end{equation}
This task requires the reservoir to model complex temporal dependencies of varying order $p$. Fig.~\ref{fig:narma_task} shows the NMSE across diffrent orders, parameterized by the fractional order $\alpha$.\\
Coherently with the observations for the linear memory capacity, the non-Markovian memory provides a significant computational advantage for short-range nonlinear processing. For NARMA orders $p \lesssim 5$, decreasing the fractional order $\alpha$ significantly reduces the prediction error compared to the strictly Markovian reservoir. The system successfully leverages its enhanced short-term memory to capture highly nonlinear dynamical dependencies. For higher orders ($p > 5$), the error accumulation increases --- directly reflecting the suppressed memory tails observed in the STM evaluation and consistent with the fading memory plateau discussed in the previous Section. Tuning the fractional order, therefore, optimizes the quantum reservoir for accurate short-to-medium range nonlinear temporal processing.
This is clearly visible in Fig.~\ref{fig:narma_task}, where the NMSE bars cross as a function of $p$. Indeed, non-Markovian reservoirs ($\alpha < 1$) outperform the Markovian baseline at low orders, while the trend inverts for large $p$, confirming that fractional dynamics redistribute computational power from long-range to short-range nonlinear processing.
The environment does not merely act as a contractive sink, but the non-Markovian dynamics re-inject past correlations into the current state, providing the necessary memory to minimize prediction errors without increasing the physical reservoir size.

\textit{Conclusions}---Fractional non-Markovian dynamics significantly impacts quantum reservoir performance, offering key computational advantages for non-linear processing tasks while highlighting trade-offs in long-term memory capacity.
We simulate this via time subordination~\cite{demir2025}, mapping complex open-system dynamics to statistical ensembles of memoryless Markovian paths, thereby providing a computationally compact surrogate for intensive formalisms such as the hierarchical equations of motion~\cite{Dan2025}.\\
Crucially, the fading memory and ESP required for universal reservoir computing~\cite{monz_uni,maass, sannia2024dissipation,nokkala2021gaussian,chen2019learning,chen2020temporal} are strictly preserved. While extreme non-Markovianity can prolong memory retention at the cost of breaking the ESP~\cite{sasaki2025hamiltoniandrivenarchitecturesnonmarkovianquantum}, our fractional model deliberately targets a stable operational phase, enhancing short-term expressivity without compromising dynamical stability. This concentrated memory yields a clear computational trade-off. Namely, short-term linear capacity and nonlinear accuracy are significantly enhanced, at the direct expense of long-range temporal retention, when compared with the performance of the Markovian reservoir. 
While the NMSE for low-order tasks ($p_{\max} = 5$) is an order of magnitude lower than for high-order tasks ($p_{\max} = 20$) across both setups, the non-Markovian reservoir systematically achieves lower overall reconstruction errors.
Therefore, environmental coupling is not merely a contractive necessity, but an active and tunable computational resource, establishing engineered non-Markovian dynamics as a highly expressive pathway for quantum temporal processing. 
While the fractional-derivative approach provides a continuous dial to tune non-Markovianity, future research could explore this framework via process tensors \cite{Keeling2026} to capture how multi-time correlations are structurally processed during sequential tasks.
A fully automated design pipeline remains an open challenge, but the framework of this work paves the way toward task- or physics-informed quantum reservoir engineering, where Nakajima-Zwanzig memory kernels~\cite{peng2026} can be matched to the temporal correlation horizon of specific computing tasks.

\textit{Acknowledgments}---
L.N. and E.P. gratefully acknowledge financial support from Eni SpA through the co-funding of the DM 117/2023 PNRR PhD grant. F.M. and E.P. acknowledge support from the PRIN-PNRR project PhysiComp (grant No. G53D23006710001).

\textit{Data Availability}---
The code and the data supporting this work are available from the corresponding author upon reasonable request.
\appendix

\bibliography{apssamp}

@article{fujii2017quantum,
  title={Harnessing disordered-ensemble quantum dynamics for machine learning},
  author={Fujii, Keisuke and Nakajima, Kohei},
  journal={Physical Review Applied},
  volume={8},
  number={2},
  pages={024030},
  year={2017},
  publisher={APS}
}

@article{maass,
title = {On the computational power of circuits of spiking neurons},
journal = {J. Comput. Syst. Sci.},
year = {2004},
author = {Wolfgang Maass and Henry Markram}
}

@article{nakajima2019quantum,
  title={Quantum machine learning in progress: macroscopic observables as quantum resources},
  author={Nakajima, Kohei and Fujii, Keisuke and Negoro, Makoto and Mitarai, Kosuke and Kitagawa, Masahiro},
  journal={Physical Review A},
  volume={100},
  number={1},
  pages={012320},
  year={2019},
  publisher={APS}
}

@article{mujal2021opportunities,
  title={Opportunities in quantum reservoir computing and extreme learning machines},
  author={Mujal, Pere and Mart{\'\i}nez-Pe{\~n}a, Rodrigo and Nokkala, Johannes and Garc{\'\i}a-Benson, Javier and Giorgi, Gian Luca and Soriano, Miguel C and Zambrini, Roberta},
  journal={Advanced Quantum Technologies},
  volume={4},
  number={8},
  pages={2100027},
  year={2021}
}

@techreport{jaeger2001echo,
  title={The ``echo state'' approach to analysing and training bounding recurrent neural networks},
  author={Jaeger, Herbert},
  year={2001},
  institution={GMD Technical Report 148}
}

@article{govia2021nonlinear,
  title={Quantum reservoir computing with a single nonlinear oscillator},
  author={Govia, Luke CG and Ribeill, Guilhem J and Rowlands, George E and Krovi, Hari R and Ohki, Thomas A},
  journal={Physical Review Research},
  volume={3},
  number={1},
  pages={013077},
  year={2021}
}

@article{rivas2014quantum,
  author    = {Rivas, {\'A}ngel and Huelga, Susana F. and Plenio, Martin B.},
  title     = {Quantum non-{M}arkovianity: characterization, quantification and detection},
  journal   = {Reports on Progress in Physics},
  volume    = {77},
  number    = {9},
  pages     = {094001},
  year      = {2014},
  publisher = {IOP Publishing},
  doi       = {10.1088/0034-4885/77/9/094001},
}

@article{cpviolation,
  title = {Entanglement and Non-Markovianity of Quantum Evolutions},
  author = {Rivas, \'Angel and Huelga, Susana F. and Plenio, Martin B.},
  journal = {Phys. Rev. Lett.},
  volume = {105},
  issue = {5},
  pages = {050403},
  numpages = {4},
  year = {2010},
  month = {Jul},
  publisher = {American Physical Society},
  doi = {10.1103/PhysRevLett.105.050403},
  url = {https://link.aps.org/doi/10.1103/PhysRevLett.105.050403}
}

@article{Vacchini2011,
  author    = {Vacchini, Bassano and Smirne, Andrea and Laine, Elsi-Mari and Piilo, Jyrki and Breuer, Heinz-Peter},
  title     = {Markovian and non-{M}arkovian dynamics in quantum and classical systems},
  journal   = {New J. Phys.},
  volume    = {13},
  pages     = {093004},
  year      = {2011},
  doi       = {10.1088/1367-2630/13/9/093004},
  eprint    = {1106.0138},
  archivePrefix = {arXiv}
}

@article{breuer2016colloquium,
  title={Colloquium: Non-Markovian dynamics in open quantum systems},
  author={Breuer, Heinz-Peter and Laine, Elsi-Mari and Piilo, Jyrki and Vacchini, Bassano},
  journal={Reviews of Modern Physics},
  volume={88},
  number={2},
  pages={021002},
  year={2016},
  doi={10.1103/RevModPhys.88.021002}
}

@article{agarwal2024,
  title={Modelling non-{M}arkovian noise in driven superconducting qubits},
  author={Agarwal, Abhishek and Lindoy, Lachlan P and Lall, Deep and Jamet, Francois and Rungger, Ivan},
  journal={Quantum Science and Technology},
  volume={9},
  number={3},
  pages={035017},
  year={2024},
  doi={10.1088/2058-9565/ad3ded}
}

@article{haase2018,
  title={Controllable Non-{M}arkovianity for a Spin Qubit in Diamond},
  author={Haase, Jan F and Vetter, Philipp J and Unden, Thomas and Smirne, Andrea 
          and Rosskopf, Joachim and Naydenov, Boris and Stacey, Alastair 
          and Jelezko, Fedor and Plenio, Martin B and Huelga, Susana F},
  journal={Physical Review Letters},
  volume={121},
  number={6},
  pages={060401},
  year={2018},
  doi={10.1103/PhysRevLett.121.060401}
}

@book{breuer2002theory,
  title={The theory of open quantum systems},
  author={Breuer, Heinz-Peter and Petruccione, Francesco},
  year={2002},
  publisher={Oxford University Press}
}

@article{mandarino,
  title = {Intermediate-times dilemma for open quantum system: Filtered approximation to the refined weak-coupling limit},
  author = {Winczewski, Marek and Mandarino, Antonio and Suarez, Gerardo and Alicki, Robert and Horodecki, Micha\l{}},
  journal = {Phys. Rev. E},
  volume = {110},
  issue = {2},
  pages = {024110},
  numpages = {28},
  year = {2024},
  month = {Aug},
  publisher = {American Physical Society},
  doi = {10.1103/PhysRevE.110.024110},
  url = {https://link.aps.org/doi/10.1103/PhysRevE.110.024110}
}

@article{devega2017,
  title={Dynamics of non-{M}arkovian open quantum systems},
  author={de Vega, In{\'e}s and Alonso, Daniel},
  journal={Reviews of Modern Physics},
  volume={89},
  number={1},
  pages={015001},
  year={2017},
  doi={10.1103/RevModPhys.89.015001}
}

@article{verstraete2009quantum,
  title={Quantum computation and quantum-state engineering driven by dissipation},
  author={Verstraete, Frank and Wolf, Michael M and Cirac, J Ignacio},
  journal={Nature Physics},
  volume={5},
  number={9},
  pages={633--636},
  year={2009}
}

@article{non_uni_prx,
  title = {Effect of Nonunital Noise on Random-Circuit Sampling},
  author = {Fefferman, Bill and Ghosh, Soumik and Gullans, Michael and Kuroiwa, Kohdai and Sharma, Kunal},
  journal = {PRX Quantum},
  volume = {5},
  issue = {3},
  year = {2024},
  month = {Jul},
  publisher = {American Physical Society}}

@misc{non-uni.barren,
      title={Noise-induced shallow circuits and absence of barren plateaus}, 
      author={Antonio Anna Mele and Armando Angrisani and Soumik Ghosh and Sumeet Khatri and Jens Eisert and Daniel Stilck Franca and Yihui Quek},
      year={2024},
      eprint={2403.13927},
      archivePrefix={arXiv},
      primaryClass={quant-ph}
}

@article{diehl2008quantum,
  title={Quantum states and phases in driven open quantum systems with cold atoms},
  author={Diehl, Sebastian and Micheli, Andrea and Kantian, Adrian and Kraus, Barbara and B{\"u}chler, Hans Peter and Zoller, Peter},
  journal={Nature Physics},
  volume={4},
  number={11},
  pages={878--883},
  year={2008}
}

@article{Metzler2000,
title = {The random walk's guide to anomalous diffusion: a fractional dynamics approach},
journal = {Physics Reports},
volume = {339},
number = {1},
pages = {1-77},
year = {2000},
issn = {0370-1573},
doi = {https://doi.org/10.1016/S0370-1573(00)00070-3},
url = {https://www.sciencedirect.com/science/article/pii/S0370157300000703},
author = {Ralf Metzler and Joseph Klafter},
}

@article{laskin2000fractional,
  title={Fractional quantum mechanics and L{\'e}vy path integrals},
  author={Laskin, Nick},
  journal={Physics Letters A},
  volume={268},
  number={4-6},
  pages={298--305},
  year={2000}
}

@article{tarasov2008fractional,
  title={Fractional Heisenberg equation},
  author={Tarasov, Vasily E},
  journal={Physics Letters A},
  volume={372},
  number={17},
  pages={2984--2988},
  year={2008}
}

@article{tarasov2021,
  title={General Non-Markovian Quantum Dynamics},
  author={Tarasov, Vasily E},
  journal={Entropy},
  volume={23},
  number={8},
  pages={1006},
  year={2021}
}

@article{iomin2009,
  title={Fractional-time quantum dynamics},
  author={Iomin, Alexander},
  journal={Physical Review E},
  volume={80},
  number={2},
  pages={022103},
  year={2009}
}

@article{wei2023,
  title={Non-Markovian dynamics of time-fractional open quantum systems},
  author={Wei, Dongmei and Liu, Hailing and others},
  journal={Chaos, Solitons \& Fractals},
  volume={174},
  pages={113816},
  year={2023}
}

@misc{sasaki2025hamiltoniandrivenarchitecturesnonmarkovianquantum,
      title={Hamiltonian-Driven Architectures for Non-Markovian Quantum Reservoir Computing}, 
      author={Daiki Sasaki and Ryosuke Koga and Taihei Kuroiwa and Yuya Ito and Chih-Chieh Chen and Tomah Sogabe},
      year={2025},
      eprint={2505.14450},
      archivePrefix={arXiv},
      primaryClass={quant-ph},
      url={https://arxiv.org/abs/2505.14450}, 
}

@misc{perego2026,
  title={Experimental characterization of coherent and non-Markovian errors using tangent space decomposition},
  author={Perego, Elia and Rodriguez-Blanco, Andrea and Whaley, K. Birgitta and Hebbe Madhusudhana, Bharath},
  year={2026},
  eprint={2602.20128},
  archivePrefix={arXiv},
  primaryClass={quant-ph},
  url={https://arxiv.org/abs/2602.20128}
}

@article{kobayashi2026edge,
  title={Edge of Many-Body Quantum Chaos in Quantum Reservoir Computing},
  author={Kobayashi, Kaito and Motome, Yukitoshi},
  journal={Physical Review Letters},
  volume={136},
  number={4},
  pages={040602},
  year={2026},
  publisher={American Physical Society},
  doi={10.1103/j2qj-vwcl}
}

@article{nakajima2020physical,
  title={Physical reservoir computing --- an introductory perspective},
  author={Nakajima, Kohei},
  journal={Japanese Journal of Applied Physics},
  volume={59},
  pages={060501},
  year={2020},
  doi={10.35848/1347-4065/ab8d4f}
}

@incollection{fujii2021chapter,
  title={Quantum Reservoir Computing: A Reservoir Approach Toward Quantum 
         Machine Learning on Near-Term Quantum Devices},
  author={Fujii, Keisuke and Nakajima, Kohei},
  booktitle={Reservoir Computing},
  series={Natural Computing Series},
  publisher={Springer},
  address={Singapore},
  year={2021},
  doi={10.1007/978-981-13-1687-6_18}
}

@article{gotting2023,
  title={Exploring quantumness in quantum reservoir computing},
  author={G{\"o}tting, Niclas and Lohof, Frederik and Gies, Christopher},
  journal={Physical Review A},
  volume={108},
  number={5},
  pages={052427},
  year={2023},
  doi={10.1103/PhysRevA.108.052427}
}

@article{kora2024,
  title={Frequency- and dissipation-dependent entanglement advantage in 
         spin-network quantum reservoir computing},
  author={Kora, Youssef and Zadeh-Haghighi, Hadi and Stewart, Terrence C 
          and Heshami, Khabat and Simon, Christoph},
  journal={Physical Review A},
  volume={110},
  pages={042416},
  year={2024},
  doi={10.1103/PhysRevA.110.042416}
}

@article{palacios2024,
  title={Role of coherence in many-body {Q}uantum {R}eservoir {C}omputing},
  author={Palacios, Ana and Mart{\'\i}nez-Pe{\~n}a, Rodrigo and Soriano, Miguel C 
          and Giorgi, Gian Luca and Zambrini, Roberta},
  journal={Communications Physics},
  volume={7},
  pages={368},
  year={2024},
  doi={10.1038/s42005-024-01859-4}
}

@article{xia2023,
  title={Configured quantum reservoir computing for multi-task machine learning},
  author={Xia, Wei and Zou, Jie and Qiu, Xiaolong and Li, Xiaopeng},
  journal={Science Bulletin},
  volume={68},
  pages={2321--2329},
  year={2023},
  doi={10.1016/j.scib.2023.08.040}
}

@article{kutvonen2020,
  title={Optimizing a quantum reservoir computer for time series prediction},
  author={Kutvonen, Aki and Fujii, Keisuke and Sagawa, Takahiro},
  journal={Scientific Reports},
  volume={10},
  pages={14687},
  year={2020},
  doi={10.1038/s41598-020-71673-9}
}

@article{martinez2020ipc,
  title={Information processing capacity of spin-based quantum reservoir 
         computing systems},
  author={Mart{\'\i}nez-Pe{\~n}a, Rodrigo and Nokkala, Johannes and Giorgi, 
          Gian Luca and Zambrini, Roberta and Soriano, Miguel C},
  journal={Cognitive Computation},
  volume={15},
  pages={1440--1451},
  year={2023},
  doi={10.1007/s12559-020-09772-y}
}

@article{xiong2025role,
  title={Role of scrambling and noise in temporal information processing with quantum systems},
  author={Xiong, Weijie and Holmes, Zo{\"e} and Angrisani, Armando and Suzuki, Yudai and Chotibut, Thiparat and Thanasilp, Supanut},
  journal={arXiv preprint arXiv:2505.10080},
  year={2025}
}

@article{sannia2024engineered,
  title={Engineered dissipation to mitigate barren plateaus},
  author={Sannia, Antonio and Tacchino, Francesco and Tavernelli, Ivano and Giorgi, Gian Luca and Zambrini, Roberta},
  journal={npj Quantum Information},
  volume={10},
  number={1},
  pages={81},
  year={2024},
  publisher={Nature Publishing Group UK London}
}

@misc{sannia2025nonmarkov,
  title={Non-Markovianity and memory enhancement in Quantum Reservoir Computing},
  author={Sannia, Antonio and Rodr{\'\i}guez, Ricard Ravell and Giorgi, Gian Luca and Zambrini, Roberta},
  journal={arXiv preprint arXiv:2505.02491},
  year={2025}
}

@article{domingo2023taking,
  title={Taking advantage of noise in quantum reservoir computing},
  author={Domingo, Laia and Carlo, G and Borondo, Florentino},
  journal={Scientific Reports},
  volume={13},
  number={1},
  pages={8790},
  year={2023},
  publisher={Nature Publishing Group UK London}
}

@article{sannia2024dissipation,
  title={Dissipation as a resource for Quantum Reservoir Computing},
  author={Sannia, Antonio and Mart{\'\i}nez-Pe{\~n}a, Rodrigo and Soriano, Miguel C and Giorgi, Gian Luca and Zambrini, Roberta},
  journal={Quantum},
  volume={8},
  pages={1291},
  year={2024},
  doi={10.22331/q-2024-03-20-1291}
}

@misc{monzani2024nonunital,
  title={Non-unital noise in a superconducting quantum computer as a computational resource for reservoir computing},
  author={Monzani, Francesco and Ricci, Emanuele and Nigro, Luca and Prati, Enrico},
  year={2024},
  eprint={2409.07886},
  archivePrefix={arXiv},
  primaryClass={quant-ph},
  url={https://arxiv.org/abs/2409.07886}
}

@article{dambre2012,
  title={Information processing capacity of dynamical systems},
  author={Dambre, Joni and Verstraeten, David and Schrauwen, Benjamin 
          and Massar, Serge},
  journal={Scientific Reports},
  volume={2},
  pages={514},
  year={2012},
  doi={10.1038/srep00514}
}

@article{ricci2025damping,
  title={Quantum reservoir computing induced by controllable damping},
  author={Ricci, Emanuele and Monzani, Francesco and Nigro, Luca and Prati, Enrico},
  journal={npj Quantum Information},
  year={2026}
}

@article{kubota2023temporal,
  title={Temporal information processing induced by quantum noise},
  author={Kubota, Tomoyuki and Suzuki, Yudai and Kobayashi, Shumpei and Tran, Quoc Hoan and Yamamoto, Naoki and Nakajima, Kohei},
  journal={Physical Review Research},
  volume={5},
  number={2},
  pages={023057},
  year={2023},
  publisher={APS}
}

@article{martinez2021dynamical,
  title={Dynamical phase transitions in quantum reservoir computing},
  author={Mart{\'\i}nez-Pe{\~n}a, Rodrigo and Giorgi, Gian Luca and Nokkala, Johannes 
          and Soriano, Miguel C and Zambrini, Roberta},
  journal={Physical Review Letters},
  volume={127},
  number={10},
  pages={100502},
  year={2021},
  doi={10.1103/PhysRevLett.127.100502}
}

@article{nokkala2021gaussian,
  title={Gaussian states of continuous-variable quantum systems provide universal 
         and versatile reservoir computing},
  author={Nokkala, Johannes and Mart{\'\i}nez-Pe{\~n}a, Rodrigo and Giorgi, Gian Luca 
          and Parigi, Valentina and Soriano, Miguel C and Zambrini, Roberta},
  journal={Communications Physics},
  volume={4},
  pages={53},
  year={2021},
  doi={10.1038/s42005-021-00556-y}
}

@article{ghosh2019quantum,
  title={Quantum reservoir processing},
  author={Ghosh, Sanjib and Opala, Andrzej and Matuszewski, Micha{\l} 
          and Paterek, Tomasz and Liew, Timothy C H},
  journal={npj Quantum Information},
  volume={5},
  pages={35},
  year={2019},
  doi={10.1038/s41534-019-0149-8}
}

@article{chen2020temporal,
  title={Temporal information processing on noisy quantum computers},
  author={Chen, Jiayin and Nurdin, Hendra I and Yamamoto, Naoki},
  journal={Physical Review Applied},
  volume={14},
  number={2},
  pages={024065},
  year={2020},
  doi={10.1103/PhysRevApplied.14.024065}
}

@article{suzuki2022natural,
  title={Natural quantum reservoir computing for temporal information processing},
  author={Suzuki, Yudai and Gao, Qi and Pradel, Ken C and Yasuoka, Kenji 
          and Yamamoto, Naoki},
  journal={Scientific Reports},
  volume={12},
  pages={1353},
  year={2022},
  doi={10.1038/s41598-022-05061-w}
}

@article{mujal2023timeseries,
  title={Time-series quantum reservoir computing with weak and projective measurements},
  author={Mujal, Pere and Mart{\'\i}nez-Pe{\~n}a, Rodrigo and Giorgi, Gian Luca 
          and Soriano, Miguel C and Zambrini, Roberta},
  journal={npj Quantum Information},
  volume={9},
  pages={16},
  year={2023},
  doi={10.1038/s41534-023-00682-z}
}

@article{maass2002liquid,
  title={Real-time computing without stable states: a new framework for neural 
         computation based on perturbations},
  author={Maass, Wolfgang and Natschl{\"a}ger, Thomas and Markram, Henry},
  journal={Neural Computation},
  volume={14},
  number={11},
  pages={2531--2560},
  year={2002},
  doi={10.1162/089976602760407955}
}

@article{chen2019learning,
  title={Learning nonlinear input--output maps with dissipative quantum systems},
  author={Chen, Jiayin and Nurdin, Hendra I},
  journal={Quantum Information Processing},
  volume={18},
  number={7},
  pages={198},
  year={2019},
  doi={10.1007/s11128-019-2311-9}
}

@article{hu2024overcoming,
  title={Overcoming the coherence time barrier in quantum machine learning 
         on temporal data},
  author={Hu, Fangjun and Khan, Saeed A and Bronn, Nicholas T and Angelatos, 
          Gerasimos and Rowlands, Graham E and Ribeill, Guilhem J 
          and T{\"u}reci, Hakan E},
  journal={Nature Communications},
  volume={15},
  pages={7491},
  year={2024},
  doi={10.1038/s41467-024-51162-7}
}

@article{boyd.chua,
  author={Boyd, S. and Chua, L.},
  journal={IEEE Transactions on Circuits and Systems}, 
  title={Fading memory and the problem of approximating nonlinear operators with {V}olterra series}, 
  year={1985},
  volume={32},
  number={11},
  pages={1150-1161}}

@article{tigher.esp,
  author={Buehner, M. and Young, P.},
  journal={IEEE Transactions on Neural Networks}, 
  title={A tighter bound for the echo state property}, 
  year={2006},
  volume={17},
  number={3},
  pages={820-824}}

@article{Stanislavsky2008,
title = {Two-time scale subordination in physical processes with long-term memory},
journal = {Annals of Physics},
volume = {323},
number = {3},
pages = {643-653},
year = {2008},
issn = {0003-4916},
doi = {https://doi.org/10.1016/j.aop.2007.04.011},
url = {https://www.sciencedirect.com/science/article/pii/S0003491607000565},
author = {Aleksander Stanislavsky and Karina Weron},
}

@book{vacchini_book,
  author    = {Vacchini, Bassano},
  title     = {Open Quantum Systems},
  subtitle  = {Foundations and Theory},
  year      = {2024},
  publisher = {Springer},
  location  = {Cham},
  series    = {Graduate Texts in Physics},
  edition   = {1},
  doi       = {10.1007/978-3-031-58218-9},
  isbn      = {978-3-031-58217-2},
  pagetotal = {420}
}

@article{narma,
  author={Atiya, A.F. and Parlos, A.G.},
  journal={IEEE Transactions on Neural Networks}, 
  title={New results on recurrent network training: unifying the algorithms and accelerating convergence}, 
  year={2000},
  volume={11}}

@article{peng2026,
    author = {Peng, Bo and Zhang, Yu},
    title = {A fractional calculus framework for open quantum dynamics: From Liouville to Lindblad to memory kernels},
    journal = {The Journal of Chemical Physics},
    volume = {164},
    number = {8},
    pages = {084103},
    year = {2026},
    month = {02},
    issn = {0021-9606},
    doi = {10.1063/5.0312309}
}

@Article{Dan2025,
author={Dan, Xiaohan
and Geva, Eitan
and Batista, Victor S.},
title={Simulating Non-Markovian Quantum Dynamics on NISQ Computers Using the Hierarchical Equations of Motion},
journal={Journal of Chemical Theory and Computation},
year={2025},
month={Feb},
day={25},
publisher={American Chemical Society},
volume={21},
number={4},
pages={1530-1546},
issn={1549-9618},
doi={10.1021/acs.jctc.4c01565},
url={https://doi.org/10.1021/acs.jctc.4c01565}
}

@misc{demir2025,
      title={fractional-time deformation of quantum coherence in open systems: a non-markovian framework beyond lindblad dynamics}, 
      author={Taylan Demir},
      year={2025},
      eprint={2512.17144},
      archivePrefix={arXiv},
      primaryClass={quant-ph},
      url={https://arxiv.org/abs/2512.17144}, 
}

@book{rc.naka,
    author = { Nakajima, Kohei and  Fischer, Ingo},
    title = {Reservoir Computing. Theory, Physical Implementations, and Applications},
    publisher = {Springer Singapore} ,
    year = {2021}}

@article{monz_uni,
  title={Universality conditions of unified classical and quantum reservoir computing},
  author={Monzani, Francesco and Prati, Enrico},
  journal={Neurocomputing},
  pages={130391},
  year={2025},
  publisher={Elsevier}
}

@article{tanaka2019recent,
  title={Recent advances in physical reservoir computing: a review},
  author={Tanaka, Gouhei and Yamane, Toshiyuki and H{\'e}roux, Jean Benoit and Nakane, Ryosho and Kanazawa, Naoki and Takeda, Seiji and Numata, Hidetoshi and Nakano, Daiju and Hirose, Akira},
  journal={Neural Networks},
  volume={115},
  year ={2019},
  publisher={Elsevier}
}

@article{grollier,
    author = {Marković, Danijela and Grollier, Julie},
    title = "{Quantum neuromorphic computing}",
    journal = {Applied Physics Letters},
    volume = {117},
    year = {2020}}

@article{senanian2024microwave,
  title={Microwave signal processing using an analog quantum reservoir computer},
  author={Senanian, Alen and Prabhu, Sridhar and Kremenetski, Vladimir and Roy, Saswata and Cao, Yingkang and Kline, Jeremy and Onodera, Tatsuhiro and Wright, Logan G and Wu, Xiaodi and Fatemi, Valla and others},
  journal={Nature Communications},
  volume={15},
  year={2024}
}

@misc{franceschetto,
      title={Harnessing quantum back-action for time-series processing}, 
      author={Giacomo Franceschetto and Marcin Płodzień and Maciej Lewenstein and Antonio Acín and Pere Mujal},
      year={2024},
      howpublished={arXiv:2411.03979},
      archivePrefix={arXiv}
}

@article{ghosh2021realising,
  title={Realising and compressing quantum circuits with quantum reservoir computing},
  author={Ghosh, Sanjib and Krisnanda, Tanjung and Paterek, Tomasz and Liew, Timothy C. H.},
  journal={Communications Physics},
  volume={4},
  year={2021},
  publisher={Nature Publishing Group}
}

@article{gotting2023exploring,
  title={Exploring quantumness in quantum reservoir computing},
  author={G{\"o}tting, Niclas and Lohof, Frederik and Gies, Christopher},
  journal={Physical Review A},
  volume={108},
  number={5},
  pages={052427},
  year={2023},
  publisher={APS}
}

@article{ph1,
  title = "Scalable reservoir computing on coherent linear photonic processor",
  author = "Nakajima, Mitsumasa and Tanaka, Kenji and Hashimoto, Toshikazu",
  journal = "Communications Physics",
  volume = "4",
  year = "2021"
}

@article{ph2,
  title = {Scalable Photonic Platform for Real-Time Quantum Reservoir Computing},
  author = {Garc\'{\i}a-Beni, Jorge and Giorgi, Gian Luca and Soriano, Miguel C. and Zambrini, Roberta},
  journal = {Phys. Rev. Appl.},
  volume = {20},
  issue = {1},
  year = {2023},
  month = {Jul},
  publisher = {American Physical Society}}

@article{neutral.atoms,
  title = {Quantum Reservoir Computing Using Arrays of {R}ydberg Atoms},
  author = {Bravo, Rodrigo Araiza and Najafi, Khadijeh and Gao, Xun and Yelin, Susanne F.},
  journal = {PRX Quantum},
  volume = {3},
  issue = {3},
  pages = {030325},
  numpages = {19},
  year = {2022},
  month = {Aug},
  publisher = {American Physical Society}}

@article{boso1,
  title = {Quantum reservoir computing with a single nonlinear oscillator},
  author = {Govia, L. C. G. and Ribeill, G. J. and Rowlands, G. E. and Krovi, H. K. and Ohki, T. A.},
  journal = {Phys. Rev. Res.},
  volume = {3},
  issue = {1},
  pages = {013077},
  year = {2021},
  month = {Jan},
  publisher = {American Physical Society}}

@article{Spagnolo2022,
  author = {Michele Spagnolo and James Morris and Stefano Piacentini and others},
  title = {Experimental photonic quantum memristor},
  journal = {Nature Photonics},
  year = {2022},
  volume = {16},
  pages = {318--323},
  received = {04 August 2021},
  accepted = {07 February 2022},
  published = {24 March 2022}
}

@article{pfeffer,
  title = {Hybrid quantum-classical reservoir computing of thermal convection flow},
  author = {Pfeffer, Philipp and Heyder, Florian and Schumacher, J\"org},
  journal = {Phys. Rev. Res.},
  volume = {4},
  year = {2022},

}

@article{magri,
  title = {Prediction of chaotic dynamics and extreme events: A recurrence-free quantum reservoir computing approach},
  author = {Ahmed, Osama and Tennie, Felix and Magri, Luca},
  journal = {Phys. Rev. Res.},
  volume = {6},
  year = {2024}
}

@article{dudas2023quantum,
  title={Quantum reservoir computing implementation on coherently coupled quantum oscillators},
  author={Dudas, Julien and Carles, Baptiste and Plouet, Erwan and Mizrahi, Frank Alice and Grollier, Julie and Markovi{\'c}, Danijela},
  journal={npj Quantum Information},
  volume={9},
  number={1},
  pages={64},
  year={2023},
  publisher={Nature Publishing Group UK London}
}

@article{yama,
  title = {Reservoir Computing Approach to Quantum State Measurement},
  author = {Angelatos, Gerasimos and Khan, Saeed A. and T\"ureci, Hakan E.},
  journal = {Phys. Rev. X},
  volume = {11},
  issue = {4},
  pages = {041062},
  numpages = {18},
  year = {2021}
}

@article{breuer2009,
  title = {Measure for the Degree of Non-Markovian Behavior of Quantum Processes in Open Systems},
  author = {Breuer, Heinz-Peter and Laine, Elsi-Mari and Piilo, Jyrki},
  journal = {Phys. Rev. Lett.},
  volume = {103},
  issue = {21},
  pages = {210401},
  numpages = {4},
  year = {2009},
  month = {Nov},
  publisher = {American Physical Society},
  doi = {10.1103/PhysRevLett.103.210401},
  url = {https://link.aps.org/doi/10.1103/PhysRevLett.103.210401}
}

@misc{bylicka2013,
      title={Non-Markovianity as a Resource for Quantum Technologies}, 
      author={Bogna Bylicka and Dariusz Chruściński and Sabrina Maniscalco},
      year={2013},
      eprint={1301.2585},
      archivePrefix={arXiv},
      primaryClass={quant-ph},
      url={https://arxiv.org/abs/1301.2585}, 
}

@article{Zambon2025,
   title={Quantum Processes as Thermodynamic Resources: The Role of Non-Markovianity},
   volume={134},
   ISSN={1079-7114},
   url={http://dx.doi.org/10.1103/PhysRevLett.134.200401},
   DOI={10.1103/physrevlett.134.200401},
   number={20},
   journal={Physical Review Letters},
   publisher={American Physical Society (APS)},
   author={Zambon, Guilherme and Adesso, Gerardo},
   year={2025},
   month=May }

@article{Porto2025,
   title={Temperature and non-Markovian parameter estimation in quantum Brownian motion},
   volume={112},
   ISSN={2469-9934},
   url={http://dx.doi.org/10.1103/c87r-22mn},
   DOI={10.1103/c87r-22mn},
   number={4},
   journal={Physical Review A},
   publisher={American Physical Society (APS)},
   author={Porto, João C. P. and Vieira, Carlos H. S. and da Paz, Irismar G. and Dieguez, Pedro R. and Marinho, Lucas S.},
   year={2025},
   month=Oct }

@misc{goswami2025,
      title={Non-Markovian noise limits for sustaining entanglement in multiparty quantum states}, 
      author={Suchetana Goswami and Ujjwal Sen},
      year={2025},
      eprint={2501.04526},
      archivePrefix={arXiv},
      primaryClass={quant-ph},
      url={https://arxiv.org/abs/2501.04526}, 
}

@article{Berk2021,
   title={Resource theories of multi-time processes: A window into quantum non-Markovianity},
   volume={5},
   ISSN={2521-327X},
   url={http://dx.doi.org/10.22331/q-2021-04-20-435},
   DOI={10.22331/q-2021-04-20-435},
   journal={Quantum},
   publisher={Verein zur Forderung des Open Access Publizierens in den Quantenwissenschaften},
   author={Berk, Graeme D. and Garner, Andrew J. P. and Yadin, Benjamin and Modi, Kavan and Pollock, Felix A.},
   year={2021},
   month=Apr, pages={435} }

@article{Berk2023,
   title={Extracting quantum dynamical resources: consumption of non-Markovianity for noise reduction},
   volume={9},
   ISSN={2056-6387},
   url={http://dx.doi.org/10.1038/s41534-023-00774-w},
   DOI={10.1038/s41534-023-00774-w},
   number={1},
   journal={npj Quantum Information},
   publisher={Springer Science and Business Media LLC},
   author={Berk, Graeme D. and Milz, Simon and Pollock, Felix A. and Modi, Kavan},
   year={2023},
   month=Oct }

@misc{laine2012,
      title={Nonlocal memory effects allow perfect teleportation with mixed states}, 
      author={Elsi-Mari Laine and Heinz-Peter Breuer and Jyrki Piilo},
      year={2012},
      eprint={1210.8266},
      archivePrefix={arXiv},
      primaryClass={quant-ph},
      url={https://arxiv.org/abs/1210.8266}, 
}

@article{Keeling2026,
  title = {Process Tensor Approaches to Non-Markovian Quantum Dynamics},
  author = {Keeling, Jonathan and Stoudenmire, E. Miles and Ba\~nuls, Mari-Carmen and Reichman, David R.},
  journal = {Phys. Rev. X},
  volume = {16},
  issue = {2},
  pages = {020502},
  numpages = {29},
  year = {2026},
  month = {Jun},
  publisher = {American Physical Society},
  doi = {10.1103/1ncg-11hz},
  url = {https://link.aps.org/doi/10.1103/1ncg-11hz}
}

\end{document}